\documentclass[twocolumn,showpacs,superscriptaddress,amsmath,aps,floatfix]{revtex4}
\usepackage{graphicx,color}
\usepackage{dcolumn}
\usepackage{bm}  

\begin{document}
\title{Nature of synchronization transitions in random networks of coupled oscillators }
\author{Jaegon Um}
\affiliation{School of Physics, Korea Institute for Advanced Study,
Seoul 130-722, Korea}
\author{Hyunsuk Hong}
\affiliation{Department of Physics and Research Institute of Physics and Chemistry, Chonbuk National University, Jeonju 561-756, Korea}
\author{Hyunggyu Park}
\affiliation{School of Physics, Korea Institute for Advanced Study,
Seoul 130-722, Korea}
\date{\today}
\begin{abstract}
We consider a system of phase oscillators with random intrinsic frequencies coupled through sparse random networks,
and investigate how the connectivity disorder affects the nature of collective synchronization transitions.
Various distribution types of intrinsic frequencies are considered: uniform, unimodal, and bimodal
distribution. We employ a heterogeneous mean-field approximation based on the annealed networks
and also perform  numerical simulations on the quenched Erd{\"o}s-R{\'e}nyi networks.
We find that the connectivity disorder drastically changes the nature of the synchronization transitions.
In particular, the quenched randomness completely wipes away the diversity of the transition nature
and only a continuous transition appears with the same mean-field exponent for all types of frequency
distributions. The physical origin of this unexpected result is discussed.
\end{abstract}

\maketitle              

In recent years, there has been an explosion of research on the critical
phenomena in complex networks~\cite{general}.
Most studies of various systems in complex networks
so far have been accomplished by means of
the heterogenous mean-field (MF) theory.
The heterogeneous  MF theory is based on the annealed networks where the links are not
fixed but fluctuate in time, i.e., at each time step, the neighbors of a
node are chosen randomly with its given degree.
However, the connectivity in many real systems is indeed {\em quenched}
one, i.e., the links are fixed permanently in time once they are formed.
Nevertheless, as the critical phenomena in networks are assumed to belong to some kind of
MF universality classes, it may be natural to believe that
fluctuations induced by the quenched connectivity disorder are irrelevant
in describing the MF-type critical phenomena except a finite shift of
the critical threshold. In fact, many cases such as the Ising model
and the contact process have proven to be applicable except intriguing
finite-size effects~\cite{cp,Boguna,ising}.
The synchronization problem with an unimodal distribution of random frequencies
also belongs to the case~\cite{Ichinomiya,Lee,Ott,Hong1,Hong2,Gomez1,Hong_Um,Hong_Ha}.

Very recently, there is a claim in the epidemic spreading model
via the so-called {\em quenched} MF analysis~\cite{Castellano}
that the quenched disorder affects the phase transition considerably and completely wipes away
the transition predicted by the annealed MF theory.
There, the unboundedness of degrees in networks is crucial to
make the critical threshold to vanish in the thermodynamic limit.
However, the validity of the quenched MF theory is controversial because
it still ignores dynamic (temporal) fluctuations which tend to make the endemic phase
unstable~\cite{Noh,Boguna2,Noh2}.

In this work, we present an example where the quenched connectivity disorder
changes its phase transition nature. We consider a system of coupled phase
oscillators in random networks and pay attention to its collective synchronization behavior.
In particular, we take into account various distribution types of random intrinsic frequencies
and analyze the system by means of the annealed MF theory and also
extensive numerical simulations on both quenched and
annealed networks. We find that the connectivity disorder preserves the synchronization transition at a finite
(but shifted) threshold, but sometimes with completely different transition nature.
For example, a discontinuous transition becomes continuous in the presence of the connectivity disorder.
We argue that this surprising result is due to a substantial change in the {\em effective}
random frequency distribution caused by the quenched connectivity.

We begin with a finite population of $N$ coupled phase oscillators on the
Erd{\"o}s-R{\'e}nyi (ER) random network~\cite{ER}.
To each vertex $j$ of the network, we associate an
oscillator whose state is described by the phase angle $\phi_{j}$ governed by
\begin{equation}
\dot{\phi}_{j} = \omega_{j} -J\sum_{l=1}^{N}a_{jl}\sin (\phi_{j} -\phi_{l}),
\label{eq:full}
\end{equation}
where $\omega_j$ represents the intrinsic frequency of the $j$th
oscillator, chosen from a given distribution $g(\omega)$. In this work, we restrict our
discussion to the symmetric distribution $g(\omega)=g(-\omega)$. The coupling constant
$J$ denotes the attractive coupling $(J>0)$, so the neighboring oscillators favor their
phase difference minimized.
The $\{a_{jl}\}$ is the adjacency matrix with its elements given by $a_{jl}=a_{lj}=1$ when
the vertices $j$ and $l$ are connected (linked), and $a_{jl}=0$ otherwise.
The degree of the vertex $i$ is defined as the number of linked vertices; $k_i=\sum_j a_{ij}$.
The links are fully independent of each other, then the degree distribution $P(k)$ for the ER
networks is given by the Poisson distribution~\cite{ER},
\begin{equation}
P(k)=\frac{\langle k \rangle ^k e^{-\langle k \rangle} }{k!},
\label{eq:poi}
\end{equation}
where $\langle k \rangle$ is the mean degree given by
$\langle k \rangle = \sum_{k}P(k)k$.

We explore the collective synchronization using the MF theory.
Let us first introduce a set of {\em local} ordering fields defined by
\begin{equation}
H_{j}e^{i\theta_{j}} \equiv  \sum_{l} a_{jl} e^{i\phi_{l}},
\label{eq:field}
\end{equation}
where  $H_{j}$ and $\theta_{j}$ denote the amplitude and the
mean phase of the local field at vertex $j$, respectively.
With this local field, Eq.~(\ref{eq:full}) is rewritten as
\begin{equation}
\dot{\phi}_{j} = \omega_{j} - J H_{j}\sin (\phi_{j} -\theta_{j}).
\label{eq:local}
\end{equation}
Following the previous studies~\cite{Ichinomiya, Lee, Ott, Hong1},
we assume that the link-to-link fluctuation in the local fields is
negligible under the random connection: If the network is well connected with
no fragmentation into local communities, it is expected that a cluster of
the entrained oscillators will influence on the dynamics of all oscillators
through a {\it global} ordering field $H e^{i\theta}$.
This MF assumption allows us to replace the term $e ^{i\phi_l}$ in the sum of
Eq.~(\ref{eq:field}) with the global field $H e^{i\theta}$ acting through
the edge connecting vertices $j$ and $l$. As all edges contribute the same,
the local field is simply given by the degree $k_j$ times the global
field. This is the key MF approximation taken
in~\cite{Ichinomiya, Lee, Ott, Hong1}.
In fact, this MF procedure is identical
to the MF theory on the annealed networks where the adjacency matrix
is approximately given by the connecting probability as
\begin{equation}
a_{jl}\approx \frac{k_j k_l}{N\langle k\rangle}~.
\label{aMF}
\end{equation}
In this scheme, the global field can be explicitly
given as
\begin{equation}\label{SCE}
He^{i\theta} \approx k_{j} H_j e^{i\theta_j}  \approx \frac{1}{N}\sum_l \frac{k_l}{\langle k\rangle} e^{i\phi_l}~.
\end{equation}

Substituting the local fields by the global field approximately in
Eq.~(\ref{eq:local}), the annealed MF equation is given by
\begin{equation}
\dot{\phi}_{j} = \omega_{j} -k_{j}JH\sin(\phi_{j} -\theta)~.
\label{eq:mf}
\end{equation}
Note that the oscillator at vertex $j$ feels the {\em effective} coupling constant $k_j J$
with the global field $H$.
The self-consistency equation for $H$, Eq.~(\ref{SCE}), then reads~\cite{Hong1}
\begin{equation}
H= \frac{1}{N} \sum_{j=1}^{N} \frac{k_{j}}{\langle k \rangle}
\sqrt{1-\left( \frac{\omega_{j}}{k_{j}JH} \right)^{2}}
\Theta\left( 1-\frac{|\omega_{j}|}{k_{j}JH} \right),
\label{eq:self1}
\end{equation}
where $\Theta(x)$ is the Heaviside step function: $\Theta(x)=1$ for
$x\geq 0$ and 0 otherwise. This implies that only the {\em entrained}
(synchronized in frequency) oscillators that have the intrinsic
frequencies $\omega_j$ restricted by $|\omega_j|<k_j JH$
contribute to the global field $H$.
Note that the connectivity disorder  is reduced to the degree
fluctuation only, in this annealed MF description.
In the continuum limit of $N \rightarrow \infty$, Eq.~(\ref{eq:self1})
can be rewritten as~\cite{Hong1}
\begin{equation}
H= \frac{1}{ \langle k \rangle } \sum_{k} P(k)k  U(kJH),
\label{self:inf}
\end{equation}
where
\begin{equation}
U(x)= \int_{-x}^{x} d\omega~ g(\omega)\sqrt{1-\left(\omega/x\right)^{2}}~.\label{U}
\end{equation}

For comparison, we revisit the {\em globally} interacting oscillator system on a complete graph (CG)
where $a_{jl}=1$ for all pairs of $(j,l)$ and $P(k)=\delta_{k,N}$  with the coupling constant rescaled as
$J\rightarrow J/N$. In this case, the above MF procedure is exact except dynamic fluctuations,
which yields
\begin{equation}
\dot{\phi}_{j} = \omega_{j} -JH\sin(\phi_{j} -\theta)~~\mbox{and}~~
H= U(JH)~.
\label{eq:self1c}
\end{equation}
In heterogeneous networks compared to the above CG case,
oscillators with higher degree $k$ tend to behave
with a stronger effective coupling constant $kJ$, thus become entrained earlier at smaller $J$. Therefore, oscillators
with the same intrinsic frequency $\omega$ will be entrained in the order of degree $k$ from above. However,
as the oscillator frequencies are also distributed randomly, there may be an intricate interplay between
the frequency and the degree distribution.

We first focus on a simple and interesting case that the intrinsic frequencies are drawn randomly from
a {\em uniform} (flat) and bounded distribution given by
\begin{equation}
g(\omega)= \left\{ \begin{array}{ll} \frac{1}{2b}, &
|\omega| \leq b, \\
0, & \mbox{otherwise}, \end{array} \right.
\label{eq:uniform}
\end{equation}
The function $U(x)$ in Eq.~(\ref{U}) then leads to
\begin{equation}
\label{flatU}
U(x)= \left\{ \begin{array}{ll}
\frac{1}{2b}\int_{-x}^{x} d\omega
\sqrt{1-( {\omega}/{x} )^{2}} =\frac{\pi }{4b} x&
\mbox{for}~x \leq b,  \\
\frac{1}{2b}\int_{-b}^{b} d\omega
\sqrt{1-( {\omega}/{x} )^{2}}&
\mbox{for}~x > b. \end{array} \right.
\end{equation}

For the globally interacting oscillators on the CG, it is easy to show from Eqs.~(\ref{eq:self1c}) and (\ref{flatU}) that
there is a jump of the order parameter $H$ with jump size $\Delta H=\pi/4$ at $J=J_c=4b/\pi$~\cite{Kuramoto1,Pazo}.
The reason for a finite jump is trivial mathematically, because there is no nonzero solution for $H$ possible when
$x=JH\le b$.
The first-order transition nature could be understood intuitively as follows:
For the case of a unimodal frequency distribution ($g''(0)<0$), oscillators with
small frequencies $\omega\approx 0$ can start to be entrained in frequency for a sufficiently strong coupling constant
$J_c$, and the frequency entrainment and phase synchronization spreads over to oscillators with higher $\omega$ continuously with increasing $J>J_c$.
This synchronization mechanism leads to a continuous transition.
However, with a broader distribution ($g''(0)\geq 0$), abundance of high-frequency oscillators hinders and thus destabilizes the attempted entrainment of small-frequency oscillators through direct interactions (links) between them. A flat distribution is a limiting case where
all oscillators wait to be entrained until
the highest-frequency ones ($\omega=\pm b$) become stabilized. Then, all oscillators suddenly become entrained
together in frequency (though their phases are still not fully ordered), which drives the first-order
discontinuous synchronization transition. In the case of a bimodal distribution with exponentially decaying tails,
not all but still a finite fraction of oscillators with frequencies in between two bimodal peaks
will be entrained together all of a sudden. Thus, again, a discontinuous transition is expected.
Therefore, the transition nature on the CG is quite sensitive to the characteristics of the frequency distribution function
$g(\omega)$, in particular, its curvature property at the entrainment frequency. Note that the entrainment frequency is
zero due to the symmetric property of $g(\omega)$.

Now we return to the ER network with the uniform frequency distribution. From Eqs.~(\ref{self:inf}) and (\ref{flatU}), it
is straightforward to derive
\begin{equation}
H=\frac{\pi  JH}{4b\langle k \rangle }\left[
\langle k^2 \rangle -  \sum_{k > b/JH}
 k^2 P(k) ~F\left(\frac{b}{kJH}\right)\right],
\label{eq:self2}
\end{equation}
where
\begin{equation}
F(x) = 1-\frac{2}{\pi} \left[~\sin^{-1} x
+x \sqrt{1- x^{2} }~\right]~,
\label{eq:fac1}
\end{equation}
where $F(x)$ is positive for $0\le x<1$. By a simple analysis, we find the incoherent solution,
$H=0$ for $J<J_c=\frac{4b\langle k\rangle}{\pi \langle k^2\rangle}$ and a
partially coherent solution (nonzero $H$) for $J>J_c$.

Near $J \gtrsim J_c$, the global field value $H$ can be evaluated by solving the equation
\begin{equation}
\epsilon \equiv \frac{J}{J_c}-1 \approx \frac{1}{\langle k^2 \rangle}
\sum_{k > b/J_c H}
k^2 P(k)  ~F\left(\frac{b}{kJ_c H}\right)~,
\label{incoherent1}
\end{equation}
where $\epsilon$ is the reduced coupling constant.
For small $H$, the summation is only over high $k$, where $P(k)$
decays exponentially fast with $k$ for the ER network, see Eq.~(\ref{eq:poi}).
Thus, only a few terms of $k\gtrsim b/(J_c H)$ are dominant in the
summation. Using the expansion of  $ F(x)$ near $x\lesssim 1$, the above
equation becomes
\begin{equation}
\epsilon\sim \frac{1}{\langle k^2 \rangle}
\sqrt{\frac{b}{J_c H}} ~P\left(\frac{b}{J_c H}\right)~,
\label{incoherent2}
\end{equation}
which leads to the logarithmic scaling as
\begin{equation}
H \approx \frac{b}{J_{c}} |\ln \epsilon | ^{-1}~.
\label{eq:log}
\end{equation}
Note that there is a {\em continuous} transition at $J=J_c$, even though
the logarithmic scaling implies a very steep increase of $H$ near $J\gtrsim J_c$. This can be
contrasted to the CG case where the discontinuous transition is found with
a big jump at $J=J_c$. Nevertheless, this is not quite surprising because
oscillators at vertices with many links in the tail part of $P(k)$ feel strong effective
interactions ($\sim kJ$) as discussed before, so become entrained much easily
even for high-frequency oscillators. Therefore, we expect a {\em hierarchical} synchronization (entrainment)
in the order of the degree $k$ from above.
If the degree is unbounded  with
an exponentially vanishing population, $P(k)\sim e^{-ck}$ for large $k$, the same
logarithmic scaling is expected by Eq.~(\ref{incoherent2}).
In the case of the CG case,
a similar hierarchical synchronization is also found for a continuous transition with a unimodal frequency distribution,
but in the order of the intrinsic frequency $|\omega|$ from below.
It is  interesting to note that the amplitude of the logarithmic scaling,
$b/J_c=\frac{\pi \langle k^2\rangle}{ 4 \langle k \rangle}$, depends only on the
network property and not on the width $b$ of the frequency distribution.

If we consider a random network with a finite upper bound for the degree ($k\le k_m$), the effective interaction is
also bounded. So we expect that oscillators even with the highest degree $k_m$ should wait until
the highest-frequency oscillators with $\omega=\pm b$ become stabilized, similar to the CG case. Thus, this
leads to a discontinuous transition at
$J_c=\frac{4b\langle k\rangle}{\pi \langle k^2\rangle}$ with jump  $\Delta H=b/(J_c k_m)$,
which can be easily derived from Eq.~(\ref{eq:self2}). In the limit of infinite $k_m$, the jump vanishes  and
a continuous transition is recovered.
As an example, in the case of the {\em regular} random network with $P(k)=\delta_{k,k_0}$,
we get a discontinuous synchronization transition at $J_c=\frac{4b}{\pi{k}_0}$ with
$\Delta H=\pi/4$.

For a unimodal frequency distribution with $g''(0)<0$, it is possible to find a nonzero solution for small $x=kJH$
for sufficiently high $J$~\cite{Hong1, Hong_Um}.
It can be easily seen by expanding $g(\omega)$ for small $\omega$, up to ${\cal O}(\omega^2)$, in Eq.~(\ref{U}).
Due to the symmetry of $g(\omega)$, the self-consistency equation, Eq.~(\ref{self:inf}), carries only odd-power terms in $H$
as
\begin{equation}
H \simeq \frac{\pi JH  }{2 \langle k \rangle} \left[g(0) \langle k^2 \rangle
- \frac{\langle k^4 \rangle }{8 } |g''(0)|(JH)^2 + \cdots~\right]
\label{self:expansion}
\end{equation}
The onset of synchronization is given by $J_c=\frac{2\langle k\rangle}{\pi g(0)\langle k^2 \rangle }$ and the
global field $H$ near $J\gtrsim J_c$ scales as
\begin{equation}
H \approx A ~\epsilon ^{1/2},
\label{amp}
\end{equation}
with $A^2=\frac{2 \pi^2 g^3(0)}{|g''(0)|} \frac{\langle k^2\rangle^3}{ \langle k^4 \rangle \langle k \rangle^2}$.
This ordinary  MF result is valid for any degree distribution $P(k)$ with finite $\langle k^4\rangle$, regardless of
the boundedness of the degree distribution. The cases with diverging $\langle k^4\rangle$ were discussed in details
with anomalous finite size scaling in our previous study~\cite{Hong1}.
With the gaussian $g(\omega)=g_G(\omega)=\frac{1}{\sqrt{2\pi}\sigma}e^{-\omega^2/2\sigma^2}$, the amplitude $A$ becomes independent of the width $\sigma$ of the distribution function,
i.e.~$A^2= \pi \frac{\langle k^2\rangle^3}{ \langle k^4 \rangle \langle k \rangle^2}~$.

For a binomial distribution with $g''(0)>0$, one needs higher-order terms in the expansion of Eq.~(\ref{self:expansion}) to
find a nonzero solution for $H$. But it is impossible to find a vanishingly small $H$ solution, so a discontinuous
transition is expected~\cite{Kuramoto1,Crawford}. In this work, we consider the double gaussian distribution,
i.e.~$g(\omega)=\frac{1}{2} [g_G(\omega-\omega_0)+g_G(\omega+\omega_0)]$ with $\omega_0>\sigma$.
On the CG, it is known that the so-called standing wave phase appears in between the incoherent and partially synchronized
phase~\cite{Kuramoto1,Crawford} when the bimodality becomes stronger (large $\omega_0/\sigma$). Recently, a more complex
dynamic feature was found on the CG~\cite{Ott2009}. It would be interesting to study how this feature may change in the annealed sparse network. In any case, the annealed MF theory with a binomial distribution on the ER network predicts
neither the ordinary continuous transition, nor the logarithmic continuous transition, but a discontinuous transition
at $J_c=\frac{2\langle k\rangle}{\pi g(0)\langle k^2 \rangle }$ with the random initial distribution of $\{\phi_j\}$.

In order to see whether all these interesting features found in the annealed MF
theory can persist in quenched networks, we perform
extensive numerical simulations on the ER networks.
The ER networks are generated for $\langle k\rangle=4$ up to the system size $N=64\ 000$.
Using Heun's method,
we integrate Eq.~(\ref{eq:full}) numerically with a discrete time step $\delta t=0.01$
up to $t=10^3$. Initial values for $\{\phi_j\}$ are chosen randomly and the data
are collected and averaged from $t=500$. We also average the data over $100-500$ realizations
of networks and intrinsic frequency distributions for each network size.
We also perform numerical integrations in the annealed ER networks, where
the adjacency matrix $a_{ij}$ is replaced by Eq.~(\ref{aMF}).

We measure the phase synchronization order parameter in the steady state, defined
as~\cite{Kuramoto1}
\begin{equation}
\Delta = \frac{1}{N}~\left|\sum_{j=1}^{N} e^{i\phi_{j}}\right|~.
\label{eq:delta}
\end{equation}
which can be rewritten in the annealed MF theory as~\cite{Hong1}
\begin{equation}
\Delta= \frac{1}{N} \sum_{j=1}^{N}
\sqrt{1-\left( \frac{\omega_{j}}{k_{j}JH} \right)^{2}}
\Theta\left( 1-\frac{|\omega_{j}|}{k_{j}JH} \right).
\label{eq:MFdelta}
\end{equation}
Comparing Eq.~(\ref{eq:delta}) to the global field $H$  in Eq.~(\ref{SCE}), it is easy to
see that $\Delta$ is proportional to $H$ for small $H$. In the annealed network,
one can easily derive the relation as $\Delta \approx \frac{\pi}{2} g(0) \langle k \rangle J_c H$
for small $H$ from Eq.~(\ref{eq:MFdelta}).
Therefore, the scaling behavior should be
identical for $\Delta$ and $H$ near the continuous transition and
the discontinuity in $H$, if any, should also appear in $\Delta$.

\begin{figure}[t]
\includegraphics*[width=\columnwidth]{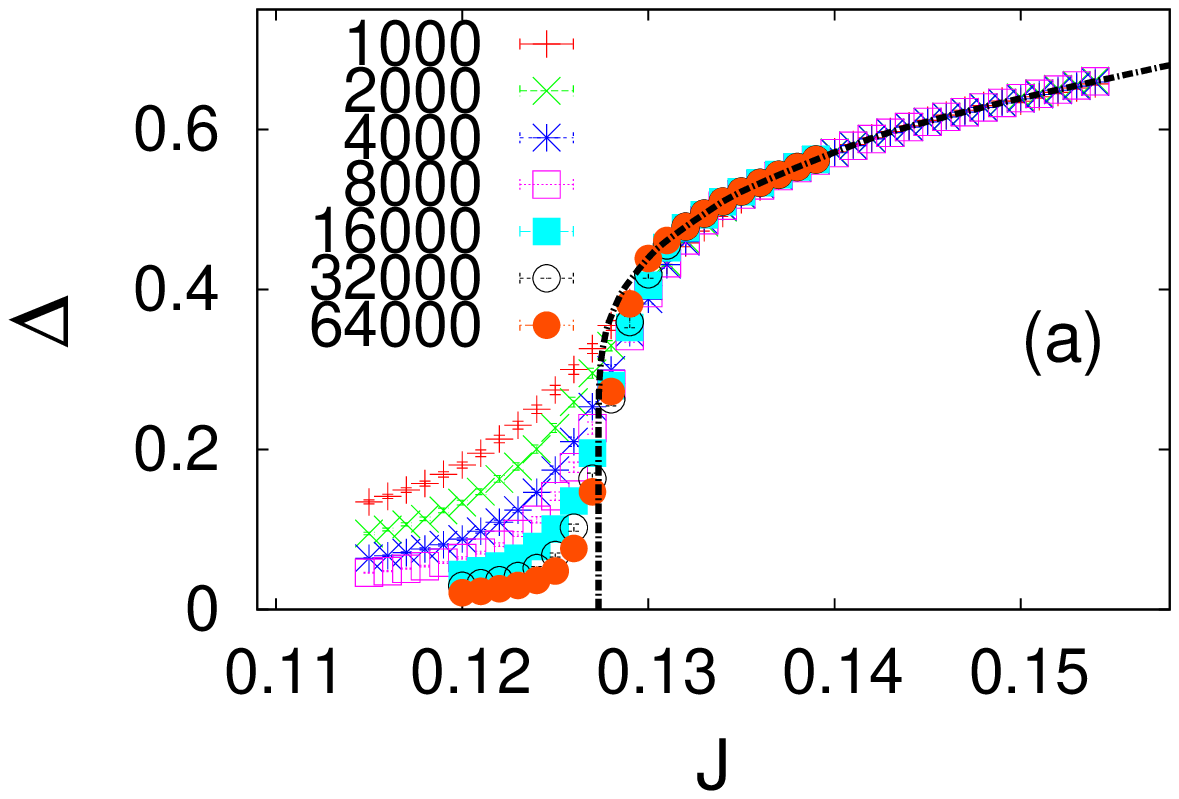}
\includegraphics*[width=\columnwidth]{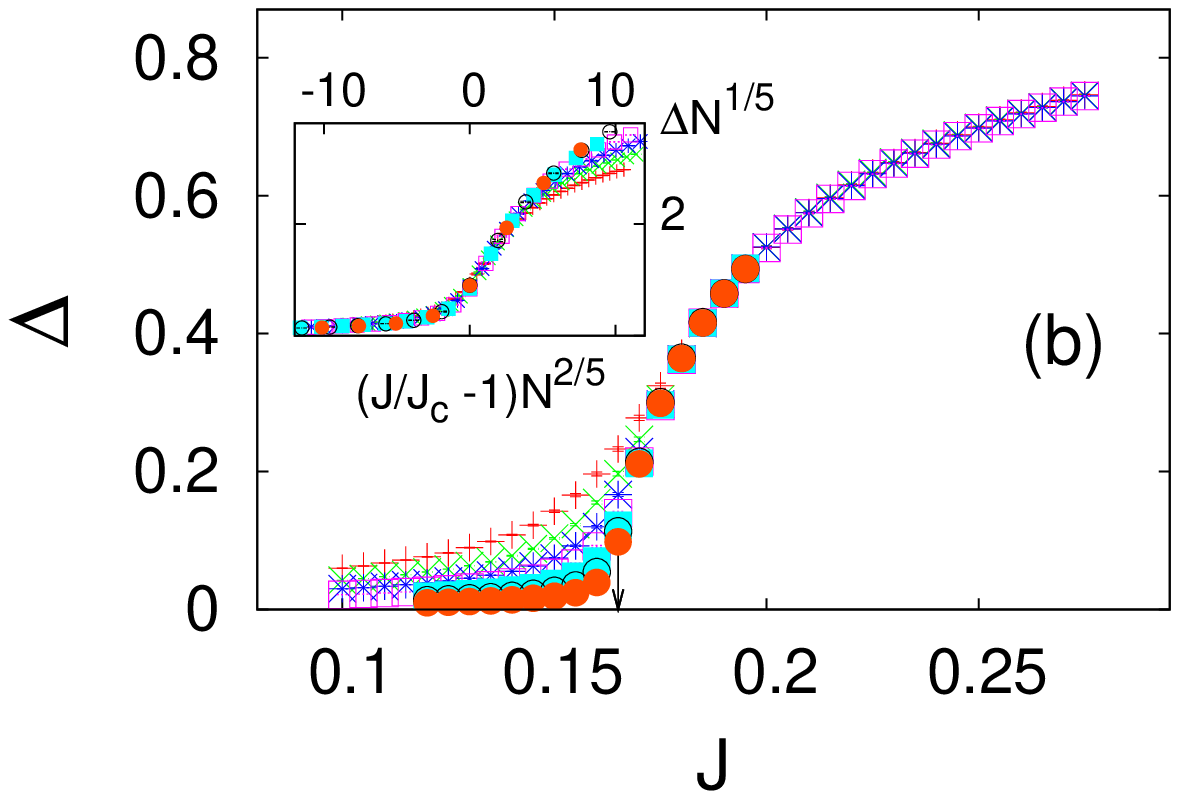}
\caption{\label{fig1} (Color online)
The order parameter $\Delta$ versus $J$ for the uniform frequency distribution with $b=1/2$
in (a) the annealed ER networks and (b) the quenched ER networks with various sizes $N$.
The dashed curve in (a) indicates the logarithmic scaling near $J_c=2/(5\pi)\approx 0.1273$ predicted
by the annealed MF theory in the $N=\infty$ limit.
Inset of (b) shows the collapse of all finite-size data on the scaling curve $f(x)$
with $\beta/\bar{\nu} =1/5$ and $\bar{\nu}=5/2$ with $J_c= 0.165(5)$ as
in Eq.~(\ref{eq:fss}).
}
\end{figure}

First, we take a uniform frequency distribution with $b=1/2$ in Eq.~(\ref{eq:uniform}).
Figure~\ref{fig1}~ shows the behavior of the order parameter
$\Delta$ as a function of the coupling strength $J$ for (a)  the annealed ER network and
(b) the quenched one. For the annealed networks, a very steep increase of
the order parameter is found for large $N$ near the exact $J_c=2/(5\pi)\approx 0.1273$, which is consistent
with the logarithmic scaling in Eq.~(\ref{eq:log}). In fact, by solving Eq.~(\ref{eq:self2}) for $H$
in the $N=\infty$ limit and evaluating $\Delta$ of Eq.~(\ref{eq:MFdelta}), the dashed curve is drawn in
Fig.~\ref{fig1}(a), which serves well as the asymptotic limit.

In contrast, for the quenched networks in Fig.~\ref{fig1}(b),
the order parameter increases rather smoothly near $J_c\approx 0.165(5)$.
To investigate the transition nature more precisely, we utilize the standard finite-size-scaling (FSS)
theory as
\begin{equation}
\Delta N^{\beta/\bar{\nu}} = f\left( \epsilon N^{1/\bar{\nu}} \right)~,
\label{eq:fss}
\end{equation}
where the scaling function $f(x)$ is given by
\begin{equation}
{f}(x)  \sim
\left\{
\begin{array}{ll}
x^{\beta} & x \gg 0, \\
\mbox {const.} & x=0, \\
(-x)^{-(\bar\nu/2-\beta)} & x \ll 0,
\end{array}
\right.\label{eq:fss}
\end{equation}
with the order parameter exponent $\beta$ and the FSS exponent $\bar{\nu}$.
This scaling behavior yields $\Delta \sim \epsilon^{\beta}$ for $\epsilon>0$ in the limit of large $N$,
$\Delta \sim N^{-\beta/\bar\nu}$ at $\epsilon=0$,
and $\Delta \sim N^{-1/2}$ for $\epsilon <0$ for large $N$.
The inset of Fig.~\ref{fig1}(b) shows an excellent agreement with the FSS with
\begin{equation}
\beta=1/2\quad \mbox{and} \quad \bar\nu=5/2~.
\label{exponent}
\end{equation}
This indicates that the synchronization transition belongs to the ordinary MF universality class with $\beta=1/2$. Moreover, the FSS exponent value of $\bar\nu=5/2$ agrees with the analytic result of the CG
case with a unimodal distribution with frequency fluctuations\cite{Hong3,Hong4}. Thus,
quite surprisingly, for the uniform distribution,
the synchronization transition nature changes from a discontinuous to a logarithmic and finally to the
ordinary MF continuous transition, as the underlying network topology changes from the CG to the annealed and finally
to the quenched ER networks.
We have also performed the numerical integrations in the quenched regular random network with $k_0=4$, and found
the same ordinary MF transition (not shown here). This implies that all distinctive features in the transition nature
are washed away when the quenched disorder in connectivity is introduced.

\begin{figure}
\includegraphics*[width=\columnwidth]{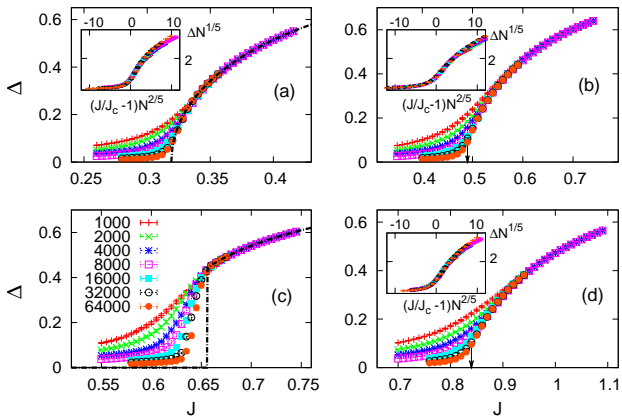}
\caption{\label{fig1b} (Color online)
The order parameter $\Delta$ versus $J$ for the gaussian (unimodal)
frequency distribution with $\sigma=1$
in (a) the annealed ER networks and (b) the quenched ER networks,
and for the double gaussian (bimodal) distribution with $\sigma=1$ and $\omega_0=1.2$
in (c) the annealed ER networks and (d) the quenched ER networks
with various sizes $N$.
The dashed curves in (a) and (c) indicate the annealed MF prediction in the $N=\infty$ limit.
Insets of (a), (b), and (d) show the collapse of all finite-size data on the scaling curve $f(x)$
with $\beta/\bar{\nu} =1/5$ and $\bar{\nu}=5/2$ with $J_c=2\sqrt{2}/(5\sqrt{\pi})\approx 0.319$,
$0.49(1)$, and $0.84(1)$,
respectively, as in Eq.~(\ref{eq:fss}). In (c), as predicted from the annealed MF theory,
the finite-size data seem to indicate the discontinuous transition at
$J_c=2\sqrt{2}/(5\sqrt{\pi})*e^{0.72}\approx 0.656$ with random initial conditions.
}
\end{figure}

Similar to the uniform distribution, we have performed the numerical
integrations for the gaussian (unimodal) frequency distribution with $\sigma=1$ and
the double gaussian (bimodal) one with $\sigma=1$ and $\omega_0=1.2$, both in
the annealed ER network and in the quenched network.
As expected, Fig.~\ref{fig1b}(a) and (b) show that  the case with the unimodal distribution
exhibits the ordinary MF continuous transition in both networks~\cite{Hong_Um}.
Also, Fig.~\ref{fig1b}(c) confirms the discontinuous transition predicted by the annealed
MF theory. However, surprisingly again, the case with the bimodal distribution exhibits the simple MF transition with $\beta=1/2$ and $\bar\nu=5/2$ in the quenched network, as seen in
Fig.~\ref{fig1b}(d). Moreover, there is no indication
of the presence of any dynamic phase including the standing wave phase. Hence, the quenched
disorder fluctuations in network connectivity seem to wipe out all interesting features found in the CG,
and drives all synchronization transitions into the ordinary MF universality class.
We also examined the extreme bimodal case with only two symmetric frequencies allowed,
i.e.~$g(\omega)=[\delta (\omega-\omega_0) + \delta (\omega+\omega_0)]/2$, and found
the same ordinary MF synchronization transition in the quenched ER networks (not shown here).

Summarizing our numerical results, the ordinary MF synchronization transition
with $\beta=1/2$ and $\bar\nu=5/2$ is found
in the quenched ER and regular networks, regardless of the intrinsic frequency distribution function $g(\omega)$. This is quite remarkable, because the transition nature
crucially depends on the shape of $g(\omega)$ in the annealed networks and also in the CG.
It obviously raises a question how the quenched connectivity disorder affects the transition
nature, against the conventional wisdom that quenched disorder fluctuations in the MF regime are irrelevant in terms of the universality.
It is also noteworthy to mention that the scaling function $f(x)$ in the ordinary MF universality class is not universal
by itself, i.e.~varies with the frequency distribution $g(\omega)$ and the underlying network structure.

\begin{figure}[t]
\includegraphics*[width=\columnwidth]{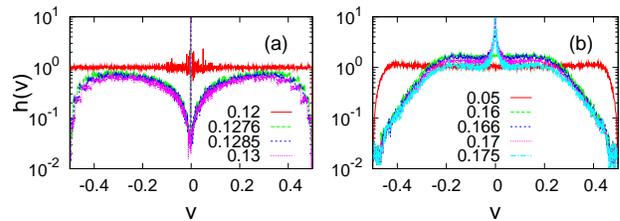}
\caption{\label{fig2} (Color online)
Histogram $h(v)$ of the mean angular velocity $v$ of oscillators in the
steady state for the uniform frequency distribution with $b=1/2$ in (a) the annealed
ER network and (b) the quenched network. The data are obtained from one sample
simulation of each network of big size $N=512\ 000$.
Note that $J_c\approx 0.1273$, and $0.165$, respectively, for each network.
}
\end{figure}

Now we explore the role of the quenched connectivity disorder in synchronization.
First, we measure the mean angular velocity of the $j$th oscillator in the steady state, defined by
\begin{equation}
v_j =  \frac{\phi_j(t_f+t_i)- \phi_j(t_i)}{t_f}
\label{eq:v}
\end{equation}
with the initial measurement time $t_i=500$ to reach the steady state and the large
duration time $t_f=10\ 000$ for a good numerical precision.
Then, we establish
the normalized histogram $h(v)$ from the mean velocity data $\{v_j\}$ of all $N$ oscillators,
which is shown in Fig.~\ref{fig2} at various values $J$ with the uniform frequency
distribution with $b=1/2$. The numerical precision of $v$ to establish $h(v)$ is given by $\Delta v=10^{-3}$.

In the annealed ER network as well as in the CG, it is straightforward to show
analytically that
the mean oscillator velocity $v_j$ does not change from its intrinsic frequency $\omega_j$
in the incoherent phase ($J < J_c\approx 0.1273$), so its distribution $h(v)$ also remains unchanged
from the intrinsic frequency distribution $g(\omega)$. Going into the partially
synchronized phase for $J> J_c$, a sharp $\delta$-function starts to develop at
$v=0$, which implies the emergence of the macroscopic entrainment of oscillators with the zero entrainment velocity, see Fig.~\ref{fig2}(a). Depletion of $h(v)$ near $v=0$ reveals
that most of entrained oscillators near $J\gtrsim J_c$ originate from those with
small intrinsic frequencies $\omega$.

On the other hand, in the quenched ER network,
the histogram $h(v)$ deviates from $g(\omega)$
considerably even in the incoherent phase ($J<J_c\approx 0.165$), which
indicates that the quenched connectivity causes the significant modification
of the oscillator velocity
even before the macroscopic entrainment emerges, see Fig.~\ref{fig2}(b).
In particular, high-frequency oscillators become quite slowed down and the fraction of small-frequency
oscillators increases before the transition. However,
the velocity distribution $h(v)$ seems still flat near $v=0$, thus this observation
by itself can not explain why the ordinary MF continuous transition should appear in the quenched network.
For $J\gtrsim J_c$, one can see a slow increase of the entrained oscillator peak at $v=0$,
which is consistent with the ordinary MF transition.

One important ingredient missing in the above study is the distinct local environment
surrounding oscillators such as the number of links (degree) and the velocities of
neighboring oscillators. For example, if an oscillator with small $\omega$ is linked to neighboring oscillators with large positive $\omega$ only, it is very difficult to
stabilize this oscillator due to the hindrance of neighboring ones. In contrast, some oscillators with large $\omega$ can join the entrainment rather easily if the neighboring oscillators try to cancel out their velocity $\omega$ in the opposite direction.
Moreover, abundance of high-frequency oscillators could not destabilize all low-frequency
oscillators because of the limited quenched connections between them.

\begin{figure}[t]
\includegraphics*[width=\columnwidth]{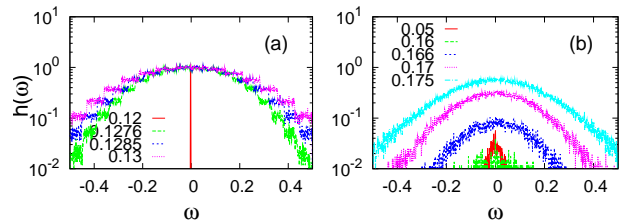}
\caption{\label{fig2_2} (Color online)
Histogram $h(\omega)$ of the intrinsic frequency $\omega$ of the entrained oscillators
for the uniform frequency distribution with $b=1/2$ in (a) the annealed
ER network and (b) the quenched network. The histogram is normalized by $N$, so its integration represents the fraction of the entrained oscillators.
The data are obtained from one sample
simulation of each network of big size $N=512\ 000$.
Note that $J_c\approx 0.1273$, and $0.165$, respectively, for each network.
}
\end{figure}

In order to see the above mechanism from the numerical simulations, we also measure
the histogram $h(\omega)$ of the intrinsic frequency $\omega$ for the entrained oscillators
at various values of $J$, see Fig.~\ref{fig2_2}. We again take the uniform frequency distribution with $b=1/2$.  As the integration of $h(\omega)$ represents the fraction of the
entrained oscillators, it is smaller than 1 for any finite $J$.
In fact, it should vanish in the $N=\infty$ limit for $J< J_c$.

In the annealed network, no oscillators with nonzero $\omega$ can be entrained for $J<J_c$,
while oscillators with  $|\omega| \leq kJH$
become entrained for $J>J_c$, which produces a step-like pattern in $h(\omega)$ due to
the discreteness of the degree distribution.
As there exists a bound of the intrinsic frequency ($|\omega|\leq b$),
all-frequency oscillators (especially including high-frequency ones) with high-enough connectivity $k$
get entrained simultaneously near the transition. As we integrate $h(\omega)$ to get
the entrainment fraction for $J\gtrsim J_c$, the significant change as a function
of $J$ occurs for large $\omega$, which again implies that the tail part of
the degree distribution $P(k)$ plays a dominant role on the critical behavior near
the transition.

In the quenched network, the situation is completely different, see Fig.~\ref{fig2_2}(b).
Some oscillators even with $\omega=0$ cannot remain entrained due to interactions
with neighboring oscillators. For $J<J_c$, the entrainment fraction is very small
around $\omega=0$ and vanishes in the $N=\infty$ limit. This vanishingly small
entrainment originates from local clusters of slow oscillators with small $\omega$
in the favorable environment
(for example, intrinsic frequencies of neighboring oscillators are
small and balanced). As $J$ approaches $J_c$ from below, these locally entrained clusters
grow in size by inviting neighboring oscillators with small $\omega$ hierarchically
and also merge together. Then, finally macroscopically entrained clusters appear
as $J$ crosses over $J_c$. As can be seen in Fig.~\ref{fig2_2}(b), the major
contribution to the entrainment comes from slow oscillators. As $J$ increases,
slow oscillators in a less favorable environment join the entrainment as well as
faster oscillators in a favorable environment. Therefore, the original shape of
$g(\omega)$, whether it is flat or bimodal, does not play an important role
near the transition. Furthermore, the shape of $h(\omega)$ near the transition
is unimodal near $\omega=0$, as slower oscillators contribute more to the entrainment.
This unimodality leads to the ordinary MF continuous transition as in the
annealed networks and also in the CG with a unimodal frequency distribution $g(\omega)$.

We can devise a simple model incorporating the quenched environment quantitatively as follows.
Neighboring oscillators in the environment affect the given oscillator by
effectively modifying its intrinsic frequency, even before the transition.
We assume that the modification
is additive and proportional to the average intrinsic frequency of neighbors,
well inside of the incoherent phase.
Then, the effective intrinsic frequency $\omega^e_j$ of the $j$th oscillator
may be written as
\begin{equation}
\omega^e_j=\omega_j +\eta_j~,
\label{mechanism}
\end{equation}
with
\begin{equation}
\eta_j \simeq \frac{\alpha (J)}{k_j} \sum_l a_{jl} \omega_l~,
\end{equation}
where $\alpha(J)$ is an unknown (increasing) function of $J$ with $\alpha(0)=0$.

For simplicity, we also assume that $\{\eta_j\}$ are independent each other
(not true if two oscillators share the same neighbor). Then,
$\langle \eta_j \eta_l \rangle=0$ for $j\neq l$. We can also get
$\langle \eta_j \rangle =0$ and
$\langle \eta_j^{2n} \rangle =\alpha^{2n} \langle \omega^{2n} \rangle /k_j^{2n-1} $
for $n=1,2,\ldots$. As the moments decrease rapidly with $n$ for $k_j>1$,
we ignore the moments for $n\ge 3$ and assume the distribution of $\eta_j$ is
gaussian, i.e. ${\tilde g}(\eta_j)=\frac{1}{\sqrt{2\pi}\tilde\sigma_j} e^{-\eta_j^2/2\tilde\sigma_j^2}$ with $\tilde\sigma_j^2=\alpha^2\langle\omega^2\rangle/k_j$.
Note that the distribution is sharper for large $k_j$.

With this distribution ${\tilde g}(\eta_j)$,
the effective frequency distribution of the oscillator $j$ is given as
\begin{eqnarray}
{\bar g}(\omega^e_j) &=& \int\int d\omega~d\eta_j~{\tilde g}(\eta_j) g(\omega)~ \delta (\omega^e_j-\omega-\eta_j) \nonumber \\
&=& \frac{1}{\sqrt{2\pi}\tilde\sigma_j} \int d\omega~e^{-{(\omega^e_j-\omega)^2}/{2\tilde\sigma_j^2}} g(\omega)~.
\end{eqnarray}
The additional concaveness in the frequency distribution near $\omega^e_j=0$ is generated due to the environmental modification $\eta_j$. The curvature ${\bar g}''(0)$ at the entrainment frequency is
\begin{equation}
{\bar g}''(0) = \frac{1}{\sqrt{2\pi}\tilde\sigma_j^3} \int d\omega~\left(\frac{\omega^2}{\tilde\sigma_j^2}-1\right)
e^{-{\omega^2}/{2\tilde\sigma_j^2}} g(\omega)~,
\label{curvature}
\end{equation}
which can be negative
with large $\tilde\sigma_j$ (low degree $k_j$ or large $\alpha$) for any distribution shape of $g(\omega)$. And this effective frequency distribution is realized well before the
macroscopic entrainment begins. So the entrainment (synchronization) mechanism operates
on the basis of the effective frequency distribution, instead of the original
intrinsic frequency distribution.
Hence, the unimodality of the effective frequency distributions of oscillators
with low degrees would be the underlying reason why the ordinary MF universality is found
for all types of $g(\omega)$.

\begin{figure}[t]
\includegraphics*[width=\columnwidth]{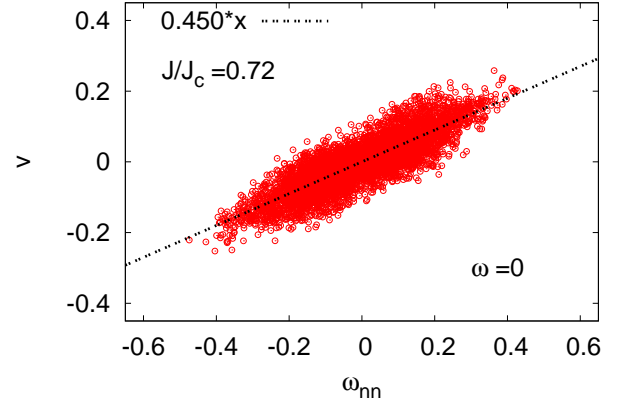}
\caption{\label{fig3_1} (Color online)
Mean velocity (effective frequency) $v$ of the $j$th oscillator versus the average intrinsic frequencies of
neighboring oscillators $\omega_{nn}=\sum_l a_{jl} \omega_l /k_j$. The mean velocity
is averaged for oscillators with $\omega_j\approx 0$ in the incoherent phase, in numerical simulations on the regular random networks of size $N=256\ 000$ with $k_0=4$ with the uniform distribution
with $b=1/2$. The dashed line is the straight line fitting the data in average.
}
\end{figure}

\begin{figure}[t]
\includegraphics*[width=\columnwidth]{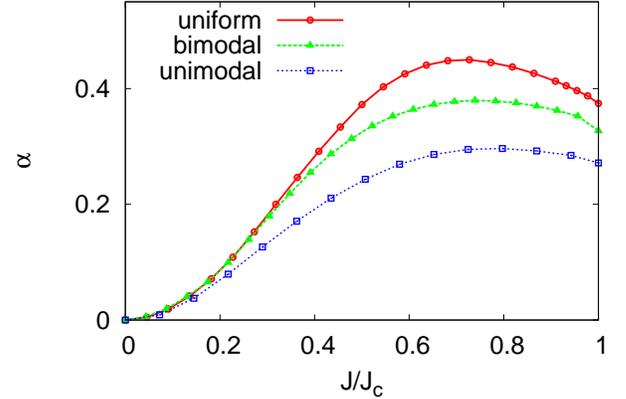}
\caption{\label{fig3_2} (Color online)
Linear slope $\alpha$ versus $J/J_c$ for various distribution types of the
intrinsic frequency $g(\omega)$.}
\end{figure}

In order to check our scenario of the effective frequency distribution, we plot
the mean velocity  (effective frequency) versus the average intrinsic frequencies of neighboring oscillators. The mean velocity is measured for oscillators with
$\omega_j\approx 0$ in numerical simulations with the uniform distribution of $g(\omega)$
with $b=1/2$ on the regular random networks with $k_0=4$. In Fig.~\ref{fig3_1}, the data
at $J=0.72 J_c$ seem to be consistent with our scenario with the linear
slope $\alpha\approx 0.45$.
We also collected data at various different values of $J<J_c$, all of which can be
fitted well with a straight line. Their linear slopes are plotted in Fig.~\ref{fig3_2}
for the unimodal, uniform, and bimodal distributions of $g(\omega)$.
As expected, $\alpha(J)$ increases at the beginning and saturates as $J$ increases.
However, it starts to decrease slightly around $70\%$ of $J/J_c$, which may be due to
the presence of local mesoscopic entrained clusters. Near $J/J_c\lesssim 1$, $\alpha$ is still
fairly finite, so the effective frequency distribution should operate well as the basis for
the emergence of macroscopic entrained clusters.

Even though the proposed simple mechanism describes qualitatively how the ordinary MF transition emerges
in quenched networks,
the quantitative prediction is still quite away from the numerical data. For example, the amplitude
of the order parameter, $A$ in Eq.~(\ref{amp}), substituting the effective distribution ${\bar g}(\omega)$,
turns out to be a few times larger than what is obtained by numerical simulations. Moreover, when the bimodality becomes bigger,
the effective curvature of Eq.~(\ref{curvature}) can remain positive, so the continuous transition is not predicted
with this simple mechanism. However, the continuous transition is found in numerical simulations.
In the extreme bimodal case with only two symmetric frequencies ($\pm\omega_0$), one can easily show that
the curvature is always positive for any $k_j$ with $\alpha<1$ and the measured value of $\alpha(J)$ in simulations
seems to be always less than $1$ as seen in Fig.~\ref{fig3_2}.

\begin{figure}[t]
\includegraphics*[width=\columnwidth]{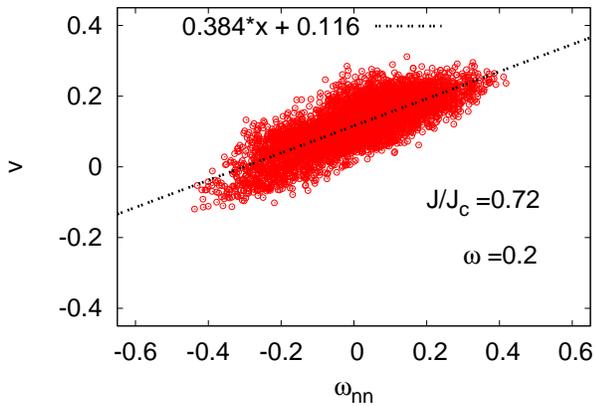}
\caption{\label{fig7} (Color online)
Mean velocity (effective frequency) $v$ of the $j$th oscillator versus the average intrinsic frequencies of
neighboring oscillators $\omega_{nn}=\sum_l a_{jl} \omega_l /k_j$. The mean velocity
is averaged for oscillators with $\omega_j\approx 0.2$ in the incoherent phase, in numerical simulations on the regular random networks of size $N=256\ 000$ with $k_0=4$ with the uniform distribution
with $b=1/2$. The dashed line is the straight line fitting the data in average.
}
\end{figure}

\begin{figure}[t]
\includegraphics*[width=\columnwidth]{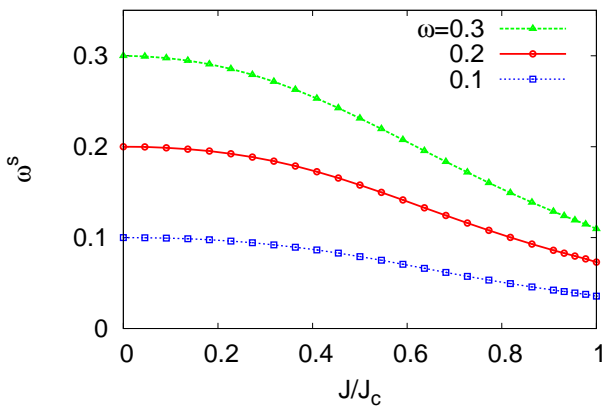}
\caption{\label{fig8} (Color online)
Shifted average velocity $\omega^s$ versus $J/J_c$ for various intrinsic frequency $\omega$'s with the uniform distribution of the
intrinsic frequency $g(\omega)$.}
\end{figure}

Therefore, there should be a secondary mechanism to fill this discrepancy. In fact, we find a tendency for
oscillators with nonzero $\omega_j$ to attain a smaller frequency in average due to the interactions with neighboring
oscillators. In Fig.~\ref{fig7}, the mean velocity of oscillators with $\omega_j\approx 0.2$ is plotted against the average intrinsic frequencies of neighboring oscillators. It shows a nice straight line but with a shifted average velocity
$\omega_j^s\approx 0.116$. With this observation, we modify our simple mechanism in Eq.~(\ref{mechanism}) as
\begin{equation}
\omega^e_j=\omega_j^s(\omega_j;J) +\eta_j~,
\label{mechanism2}
\end{equation}
where $\omega_j^s(\omega_j;J)$ is the shifted average velocity with $\omega_j^s(\omega_j;0)=\omega_j$.
We plot the average shift velocity $\omega^s(J)$ in Fig.~\ref{fig8} for the uniform distribution of $g(\omega)$.
For different types of $g(\omega)$, a similar behavior is found.
One can notice that the shift is quite sizable as $J$ approaches $J_c$,
especially for large $\omega$. This strong shift towards a smaller frequency should strengthen the unimodality of the
effective frequency distribution ${\bar g}(\omega^e)$, which can explain the finding of the ordinary MF continuous
transition even for the extreme bimodal case in quenched networks.

In summary, we have investigated the synchronization transition
of the random frequency oscillators coupled through sparse random networks.
In particular, we considered various different shapes of intrinsic frequency distributions
in the ER networks, and analyzed the system by means of the annealed MF theory.
The annealed MF theory predicts distinctive transition nature depending on
the curvature shape of the intrinsic frequency distribution.
However the numerical simulations in the quenched
network show the ordinary MF continuous transition with the same critical
exponents, regardless of the frequency distribution shape.
This implies that  the quenched connectivity drastically changes the nature of the synchronization
transitions. We discuss the underlying physical origin for this remarkable result
and provide various evidences how the quenched disorder changes effectively
the frequency distribution in the incoherent phase.

This research was supported by the NRF Grant No. 2012R1A1A2003678 (H.H.),
2013R1A6A3A03028463(J.U.), and 2013R1A1A2A10009722(H.P.).


%
\end{document}